\documentclass[12pt]{article}          % \documentclass[12pt,onecolumn]{article}
\usepackage{graphicx,epsf,subfigure,pstricks,pst-node,psfrag,amsthm,amssymb,amsmath,url,inputenc}
\begin{document}
\title{\huge Maximum Likelihood Ridge Regression}
\author{Robert L. Obenchain \\ Risk Benefit Statistics, Clayton, CA 94517}
\date{July 2022}
\maketitle

\begin{abstract}
{\noindent My first paper exclusively about ``ridge regression'' was published in \textit{Technometrics} and chosen
for invited presentation at the 1975 Joint Statistical Meetings in Atlanta. Unfortunately, that paper contained
a wide range of assorted details and results. Luckily, Gary McDonald's published discussion of that paper focused
primarily on my use of \textit{Maximum Likelihood estimation under normal distribution-theory}. In this review of
some results from all four of my ridge publications between 1975 and 2022, I highlight the Maximum Likelihood findings
that appear to be most important in practical application of shrinkage in regression.}
\end{abstract}

\begin{center}
****************************** \linebreak
Hoerl(1962) wrote ``A maximum likelihood solution for the ridge analysis \linebreak
has not yet been theoretically derived.'' This paper is dedicated to  \linebreak
Arthur E. Hoerl (1921-1994) and Robert W. Kennard (1923-2011) \linebreak
in memory of their pioneering spirits.
\end{center}

\section{Introduction}

The ``Ordinary Least Squares'' (OLS) estimator of the $\beta-$coefficient vector in linear regression is
$\hat{\beta}^o = (X'X)^+X'y$, where the superscript $+$ denotes the Moore–Penrose pseudo-inverse of $X'X$.
This OLS estimator is unquestionably the most ``well-known'' estimator that achieves ``maximum likelihood''
under normal distribution-theory. In fact, Rao (1973), Section \textbf{4g}, pages $264-265$, states that
$\hat{\beta}^o$ is the ``BLUE'' or ``Best Linear Unbiased'' estimator for linear models with independent
and homoscedastic error-terms because it achieves minimum MSE risk under those assumptions.

\begin{figure}
\center{\includegraphics[width=3.5in]{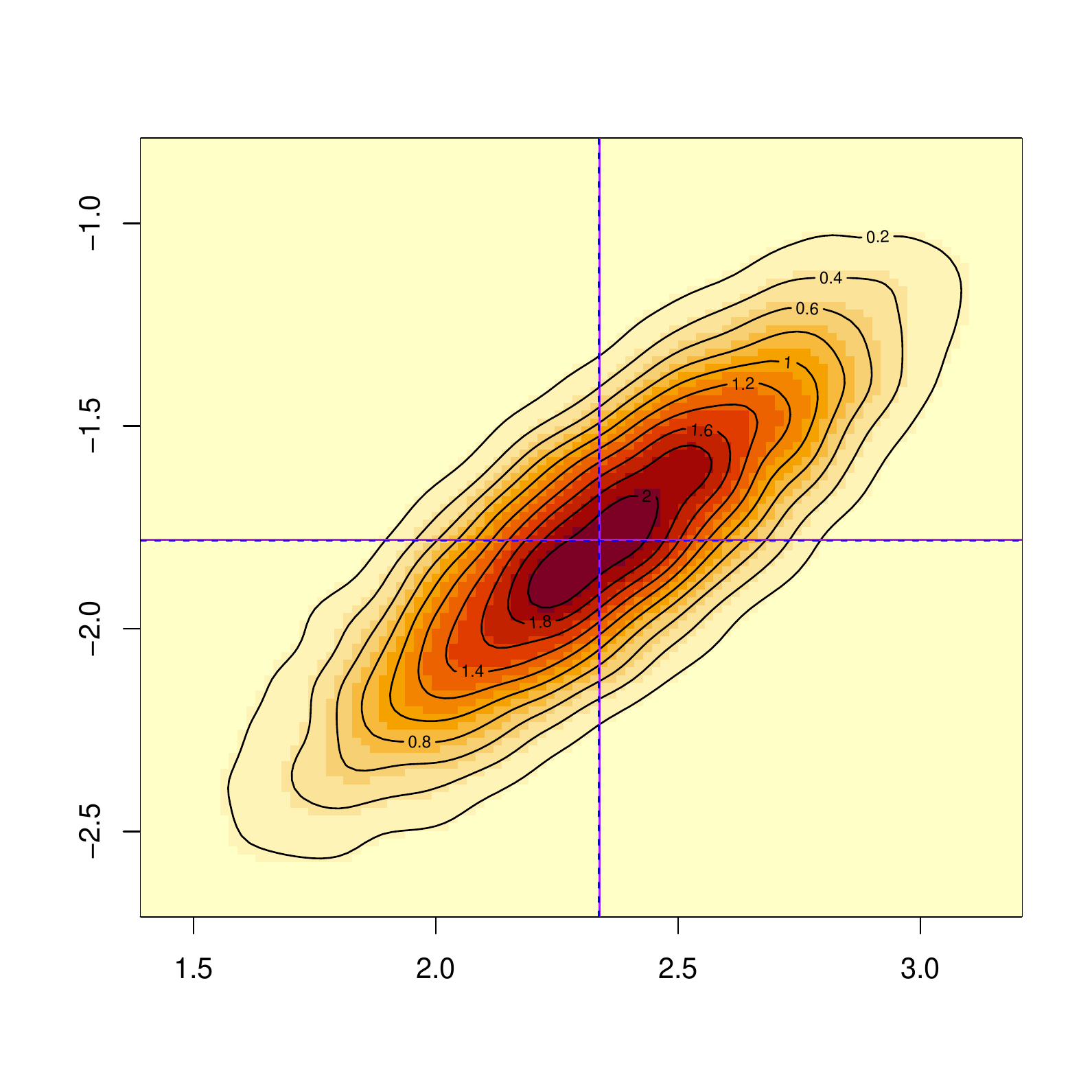}}\\
\vspace{-1.0in}
\center{\includegraphics[width=3.5in]{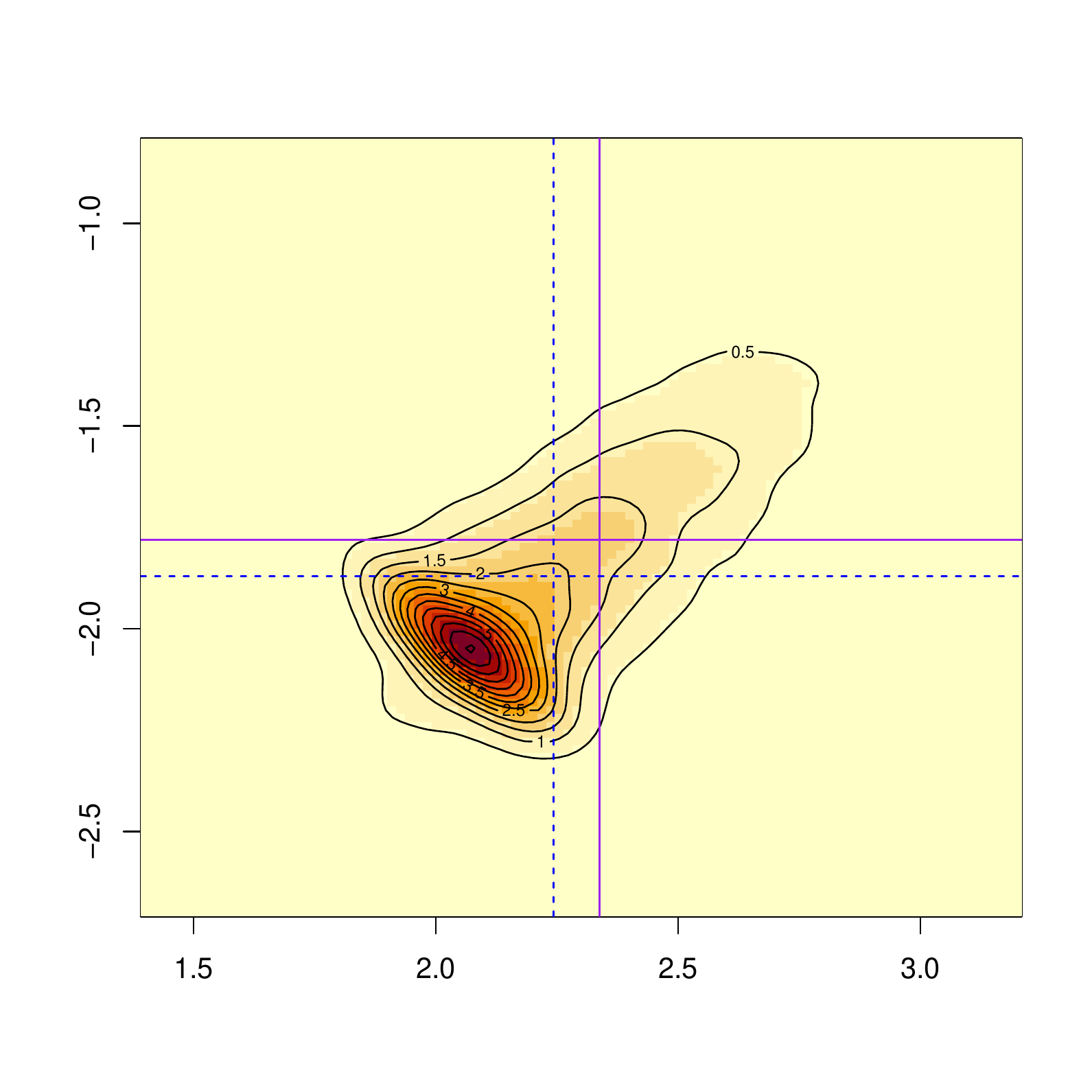}}
\caption{\label{fig:MLBB} This figure from Obenchain (2022a) displays Kernel Density estimates for a
pair of bootstrap distributions for two $\hat{\beta}-$coefficients. True coefficients $\beta_1 = 2.3382$
and $\beta_2 = -1.7809$ were used to generate $n = 500$ observations on a $y-$variable with errors that
were I.I.D. Normal with mean $0$ and $\sigma^2 = 18$. In the top plot, note that the distribution of OLS
estimates definitely appears to be unbiased but has high variability. In the bottom plot, the corresponding
distribution of GRR estimates \textit{most likely to be optimally biased} is much less variable and achieves
lower overall MSE risk.}
\end{figure}

Surprisingly, the ``Main Difficulty'' with OLS in practical applications of linear regression is that it is
unbiased: $E(\hat{\beta}^o) = \beta$. Unbiasedness implies that the variances of and the covariances between
$\hat{\beta}^o-$estimates can be quite large when the $X-$matrix of predictors has $p \geq 2$ columns that
are highly inter-correlated (ill-conditioned). See Figure (\ref{fig:MLBB}) on page $-3-$ for a simulated
illustration of this with $p = 2$ coefficients.

The well-known ridge estimators of Hoerl and Kennard (1970a,b) use a single (scalar) parameter, $k$, commonly
characterized as being a ``small positive numerical constant'' that is added to each diagonal element of the
$X'X-$matrix before it is inverted: $\hat{\beta}(k) = [X'X + k{\cdot}I]^{-1}X'y$. These $\hat{\beta}(k)$
estimators are \textit{biased} when $k > 0$, but the normal distribution-theory ``likelihood'' of these
estimators to be \textit{optimally biased} had not been quantified before the publication of Obenchain(1975).

Every ``generalized'' ridge regression shrinkage ``path''  starts at $\hat{\beta}^o$, the OLS Best estimator.
Since the \textit{terminus} of the shrinkage path is (usually) taken to be $\hat{\beta} = 0$, and the overall
length of the shrinkage path is always finite, it is somewhat unfortunate that Hoerl and Kennard's $k-$parameter
must approach $+\infty$ to actually approach this terminus. My (1975$-$1977) generalized ridge estimators
used two strictly \textit{finite parameters}, $q$ and $m$, to determine both the $q-$Shape or ``curvature'' of
a ``shrinkage path'' [$-5 \leq q \leq +5$] as well as the $m-$Extent of shrinkage [$0 \leq m \leq p$] along
that path.

Gunst (2000, p. 62) wrote: ``Although ridge regression is widely used in the application of regression methods
today, it remains as controversial as when it was first introduced.'' Many early proposals for choosing the
$k-$factor appear to have been based mainly upon heuristics. In his discussion of Obenchain (1975), McDonald (1975)
echoed the ``likelihood'' remark of Hoerl (1962) about why shrinkage heuristics or Cross-Validation or Bootstrapping
methods were commonly used to determine the $k-$factor. 

While my first three ridge publications (1975, 1977, 1978) provided a sound, theoretical ``likelihood'' foundation
for shrinkage, they focused on ``parsimonious'' $2-$parameter paths that typically miss the global maximum likelihood
$\beta-$estimator when the ``centered'' $X-$matrix has rank $> 2$. My fourth ridge publication, Obenchain (2022a),
proposed an \textit{Efficient Shrinkage Path} which usually applies a different $\delta-$factor to each ``uncorrelated component'' of the OLS estimator, $\hat{\beta}^o$. This $p-$Parameter path consists of two-piece linear functions that
first connect each (unbiased) OLS coefficient estimate to it's optimally biased estimate, and then heads directly for
the shrinkage terminus.

This (unpublished) paper outlines and updates the technical material needed by researchers to understand the details
of my published work on shrinkage in regression and, hopefully, to make genuinely ``new'' contributions to regression
methodology and/or software development.

\subsection{Multiple Regression Notation and Standardization}

The usual model for multiple linear regression is written as

\begin{equation}
E(y|X)=1\mu +X\beta \text{ and }Var(y|X)={\sigma}^2I\text{ ,}
\label{REGMOD}
\end{equation} % #1

\noindent where $y$ is a $n{\times}1$ vector of observed response values,
$X$ is a $n{\times}p$ matrix of non-constant coordinates for $p$ predictor variables,
$1$ is a $n{\times}1$ vector of ones, $\mu$ is an unknown intercept, and
$\beta$ is a $p{\times}1$ vector of unknown regression coefficients.
Note that the $X-$matrix is assumed given, while the $y-$vector of response
outcomes is assumed to be (conditionally) stochastic in the sense that it
consists of uncorrelated observations with constant unknown variance, $\sigma^2.$

In this paper, the $y-$vector and the columns of the $X-$matrix are assumed to have
been ``centered'' by subtracting off column means; thus, $1^{\prime}y=0$ and
$1^{\prime}X=0^{\prime }$. We also assume that the centered $X-$matrix has full
column rank, $p \leq (n-1)$, so that the OLS estimator of the $\beta-$vector is well-defined and
uniquely determined. Note that this centering allows model (\ref{REGMOD}) to be written
more succinctly as $E(y|X)=X\beta $ and $Var(y|X)=\sigma ^2(I-11^{\prime }/n)$, where the
$\mu$ intercept term from equation (\ref{REGMOD}) will be implicitly estimated by
$\bar y-\bar x^{\prime }\hat \beta $. A key advantage of this centering
is that the $\hat \mu $ estimate will then adjust appropriately in response
to any changes in $\hat{\beta}-$estimates. Specifically, the fitted regression
hyperplane will always pass through $y = \bar y$ at $x = \bar x$. 

Finally, we remark that a regression model can truly be ``ill-conditioned'' only when
$p > 1$. However, shrinkage can still reduce the MSE risk of a scalar $\hat{\beta}$ estimate.
The YonX() function in the \textit{RXshrink} R-package, Obenchain (2022b), provides Trace
displays and other pedagogical graphics for this (simple) $p = 1$ special-case.

\subsection{Principal Axis Rotation to Uncorrelated Components}

The singular value decomposition of regressors is $X=H\Lambda
^{1/2}G^{\prime }$, where $H$ is the $n\times p$ semi-orthogonal matrix of 
\textit{principal coordinates} of $X$, $\Lambda$ is the $p\times{p}$
diagonal matrix of \textit{eigenvalues} of $X^\prime{X}=G\Lambda{G^\prime}$,
and $G$ is the $p\times{p}$ orthogonal matrix of \textit{principal
axis direction cosines}. Note that $1^{\prime}H=0^\prime$ because
$1^{\prime}X=0^{\prime}$. Furthermore, $G^{\prime }G$ and $H^{\prime }H$
are $p\times{p}$ identity matrices. Principal axes are assumed here to be
ordered such that the eigenvalues, $\lambda_1\geq \cdots \geq \lambda_p>0$,
of the regressor inner products matrix are non-increasing.

The ordinary least-squares (OLS) estimator, $\hat{\beta}^o$, of $\beta$ in equation
(\ref{REGMOD}) is not uniquely determined when $rank(X) < p$, so we adopt the
convention here that

\begin{equation}
\hat{\beta}^o \equiv X^{+}y=Gc,  \label{OLS}
\end{equation}  % #2

\noindent where $X^{+}$ is the (unique) Moore-Penrose inverse of $X$, and $c$
denotes the $p\times 1$ vector containing the (sample) \textbf{uncorrelated components}
of $\hat{\beta}^o$, given by
 
\begin{equation}
c \equiv \Lambda^{-1/2}H^{\prime}y = (y^{\prime}y)^{1/2}\Lambda^{-1/2}\rho\text{ .} \label{UCORCOMP}
\end{equation}  % #3

\noindent In this third equation, $\rho =H^{\prime}y/\sqrt{y^{\prime}y}$ denotes the computed
$p{\times}1$ vector of \textit{observed principal correlations} between $y$ and the columns
of $H$. Since these principal-correlation estimates will appear in many equations below,
I apologize for any ambiguity caused by using a Greek letter (without a ``hat'') to denote them!

Note that $E[c|X]=G^{\prime}\beta \equiv \gamma $, which is the $p\times 1$ vector of
unknown\textit{\ true components} of $\beta $, and that $Var[c|X]=$ $\sigma ^2\Lambda ^{-1}$,
which is a diagonal $p\times{p}$ matrix.

The F-ratio for testing $\gamma _i=0$ is then

\begin{equation}
F_i=\frac{c_i^2}{s^2/\lambda _i}=\frac{(n-p-1)\cdot \rho_i^{2}}{(1-R^2)}\text{ .}
\label{FRAT}
\end{equation}   % #4

\noindent In equation (\ref{FRAT}), $R^2$ is the familiar R-squared ``Coefficient of
Determination'' that can also be written as the sum-of-squares of observed principal
correlations: $R^2=\rho_1^{2}+\cdots +\rho_p^2$. Note also that $(n-p-1)$ is the number of
``degrees-of-freedom for error'', and that $s^2 = y^{\prime}(I-HH^{\prime})y/(n-p-1)
=(y^{\prime }y)\cdot (1-R^2)/(n-p-1)$ is the unbiased estimator of $\sigma ^2$ under normal
distribution-theory.

Note further that equation (\ref{FRAT}) illustrates that the statistical
significance of the uncorrelated components depends only upon their
corresponding principal correlations. On the other hand, equation (\ref{UCORCOMP})
shows that

\begin{equation}
c_i=\rho_i\cdot \sqrt{\frac{y^{\prime }y}{\lambda _i}}\text{.}  \label{UCC2}
\end{equation}  % #5

\noindent \textbf{Key Insight} from equation (\ref{UCC2}): An uncorrelated component can
be relatively large, numerically, simply because its corresponding eigenvalue, ${\lambda _i}$,
is relatively small, even when its sample principal correlation, ${\rho_i}$, is
\textit{not relatively large}. This ``ill-conditioning'' occurs when correlations between
$X$-variables cause the centered $X'X$ matrix to have a wide range of unequal eigenvalues.
Furthermore, components with relatively small eigenvalues necessarily have relatively large
variance, $V[c_i|X]=\sigma ^2/\lambda _i$, and thus are primary candidates for generalized
ridge shrinkage.

Finally, Generalized Ridge Regression (GRR) estimators are of the form

\begin{equation}
\hat{\beta}(\Delta) \equiv G\Delta{c},  \label{BGRR}
\end{equation}   % #6

\noindent where $\Delta$ is a $p\times{p}$ \textit{diagonal matrix} containing $p$ ``shrinkage''
$\delta_i-$factors such that $0 \leq \delta_i \leq 1$ and are ideally \textit{known constants}
given the $X-$matrix of ``standardized'' predictor co-ordinates.

Equation (\ref{BGRR}) shows that Generalized Ridge Regression may possibly best be viewed as
the ``shrinkage'' version of \textit{Principal Components Regression}, Massy(1965), where each
individual $\delta_i-$factor is taken to be either $1$ or $0$.

\begin{figure}
\center{\includegraphics[width=0.5\textwidth]{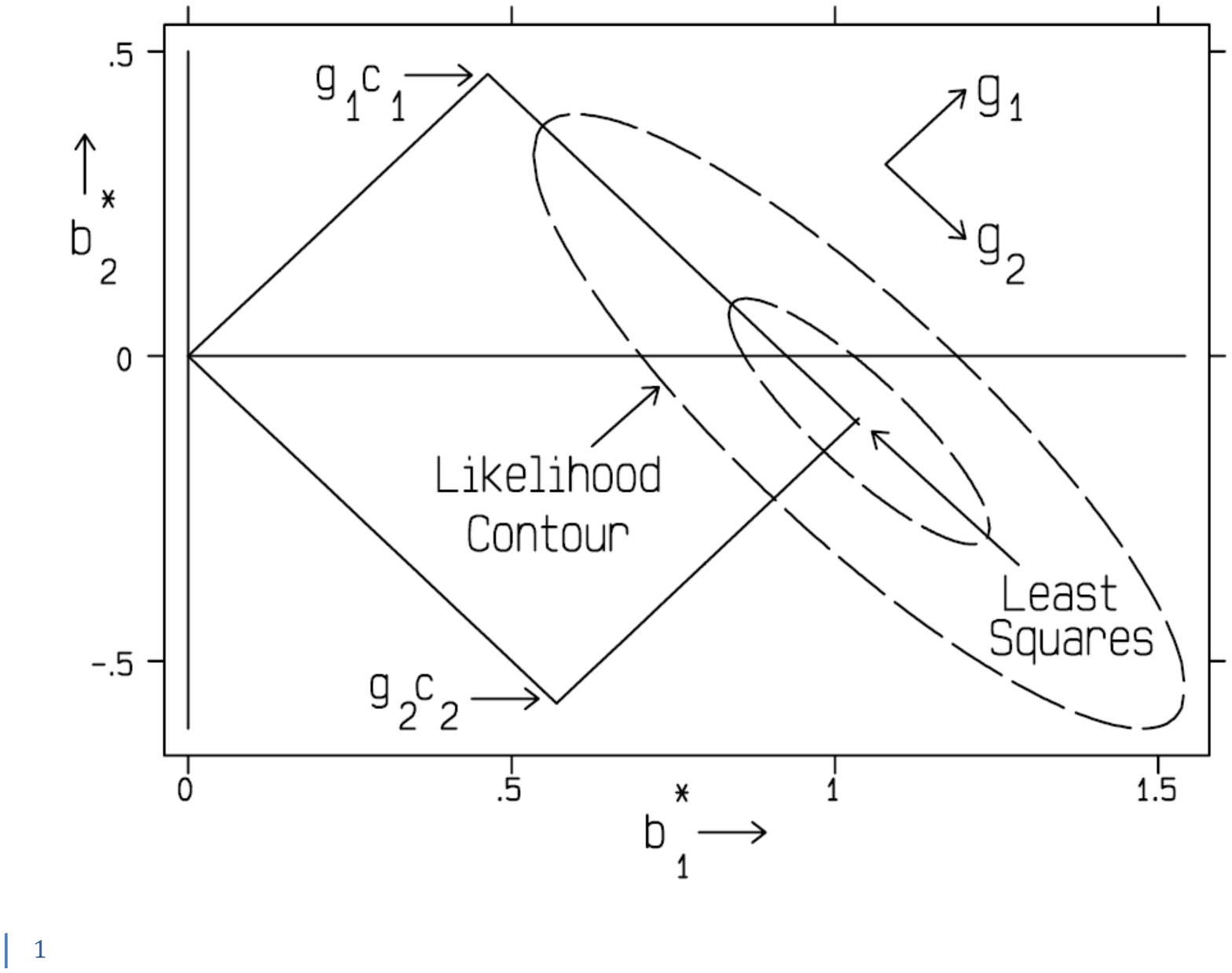}}
\caption{\label{fig:ridge1} When the ``standardized'' $X-$matrix is of rank $p=2$,
the $b^*=\hat{\beta}^o=Gc$ decomposition of equations (2) and (3) becomes
$\hat{\beta}^o=g_1c_1+g_2c_2$ where $g_1$ and $g_2$ are the columns of $G$. This principal
axis rotation (or reflection) orients the edges of the \textit{rectangle} displayed above. It
contains all shrinkage estimators of the form
$\hat{\beta}(\Delta)=G{\Delta}c=g_1\delta_1c_1+g_2\delta_2c_2$
where $0 \leq \delta_1$,$\delta_2 \leq 1$.}
\end{figure}

\subsection{Optimal Shrinkage Factors}

To apply normal-theory maximum likelihood to shrinkage estimation, we will need a
\textit{definition} for the $\Delta$ factors that make $\hat{\beta}(\Delta)=G{\Delta}c$ the
minimum MSE, linear estimator of $\beta$. We start by computing the risk of ${\delta}_i$ times
$c_i$ as an estimator of the $i-$th true component ${\gamma}_i$:

\begin{equation}
MSE(\delta _ic_i)=E[(\delta _ic_i-\gamma _i)^2]=E\{[\delta _i(c_i-\gamma_i)-(1-\delta_i)\gamma_i]^2\}.
\label{MSE1}
\end{equation}   % #7

Under the assumption that $\delta _i$ is nonstochastic given $X$, this risk expression can first
be rewritten as
\begin{equation}
MSE(\delta _ic_i)=\delta _i^2E[(c_i-\gamma _i)^2]-2\delta _i(1-\delta
_i)E(c_i-\gamma _i)+(1-\delta _i)^2\gamma _i{}^2,
\label{MSE2}   % #8
\end{equation}
\noindent and then simplified, using $E(c_i)=\gamma _i$ and
$V(c_i)=\sigma ^2/\lambda _i$, to yield:
\begin{equation}
MSE(\delta _ic_i)=\delta _i^2\sigma ^2/\lambda _i+(1-\delta _i)^2\gamma
_i{}^2.
\label{MSE3}   % #9
\end{equation}
Next, we compute the partial derivatives of MSE risk with respect to changes
in the $i-$th shrinkage factor:
\begin{equation}
\partial MSE(\delta _ic_i)/\partial \delta _i=2\delta _i\sigma ^2/\lambda
_i-2(1-\delta _i)\gamma _i{}^2,\text{ and}  \label{1deriv}
\end{equation}
\begin{equation}
\partial ^2MSE(\delta _ic_i)/\partial \delta _i^2=2\sigma ^2/\lambda
_i+2\gamma _i{}^2.  \label{2deriv}
\end{equation}   % #10
Since the second derivative will be strictly positive whenever $\sigma >0$,
the solution of $\partial MSE/\partial \delta _i=0$ corresponds to minimum
risk. This solution will be denoted by $\delta _i=\delta _i^{MSE}$, where
$\delta _i^{MSE}$ can be written in at least three equivalent ways:

\begin{equation}
\delta _i^{MSE}\equiv \frac{\gamma _i^2}{\gamma _i^2+(\sigma ^2/\lambda _i)}=\frac{%
\lambda _i}{\lambda _i+(\sigma ^2/\gamma _i^2)}=\frac{\varphi _i^2}{\varphi _i^2+1}
\text{ ,} \label{DOPT}
\end{equation}   % #11

\noindent where $\varphi _i^2=\gamma _i^2\lambda _i/\sigma ^2$ is the (unknown)
noncentrality of the F-ratio for testing $\gamma _i=0,$ equation (\ref{FRAT}).
In fact, $n\cdot F_i/(n-r-1)$ is the maximum likelihood estimator of
the $\varphi _i^2$ noncentrality parameter under normal distribution-theory.

Note that $0\leq \delta _i^{MSE}\leq 1$. Furthermore, $\delta _i^{MSE}=0$
only when $\gamma _i=0$ or in the limit as $\sigma ^2$ increases to $+\infty $.
Similarly, $\delta _i^{MSE}=1$ only when $\sigma ^2=0$ or in the limit as 
$|\gamma _i|$ increases to $+\infty $.

Substituting $\delta _i^{MSE}$ into equation (\ref{MSE1}) and simplifying, we
have shown that

\begin{equation}
MSE(\delta _ic_i)\geq \delta _i^{MSE}\cdot (\sigma ^2/\lambda _i  \label{MSE4})
\end{equation}   % #12
for all non-stochastic shrinkage factors $0\leq \delta _i\leq 1$, with
equality only when $\delta _i=\delta _i^{MSE}$. 

One never really knows when a non-stochastic choice for $\delta _i$ is equal
to $\delta _i^{MSE}$; after all, $\delta _i^{MSE}c_i=c_i/[1+(\sigma^2/\lambda _i
\gamma _i^2)]$ is a non-estimator of $\gamma _i$ because its $(\sigma
^2/\gamma _i^2)$ factor is unknown. The OLS estimator, $c_i$, is
the unique minimax estimator of $\gamma _i$ under normal distribution-theory;
its risk, $MSE(c_i)=\sigma ^2/\lambda _i$, is constant for all
values of $\gamma _i$.

Obenchain(1978) considered alternative definitions for MSE optimal
shrinkage but ultimately concluded that (\ref{DOPT}) is the most reasonable
definition overall.

\section{Likelihood in Shrinkage Estimation}

Equation (\ref{DOPT}) can be inverted to yield
\begin{equation}
\gamma_{i}^{2}/\sigma^2 = \delta_{i}^{MSE}/[\lambda_{i}(1 - \delta_{i}^{MSE})]\text{ .}  \label{KeyEqn}
\end{equation}    % #13
\noindent Equation (\ref{KeyEqn}) quantifies the KEY relationships used in Obenchain (1975) to
define the normal distribution-theory likelihood that any given set of $p$ shrinkage
$\delta-$factors achieves overall minimum MSE-risk.

Specifically, the likelihood that any given set of numerical shrinkage $\delta-$factors within the
half-open interval [$0 \leq \delta_{i} < 1$) are MSE-optimal is defined to equal the likelihood that
$\gamma_{i}$ is $\pm \sigma^{**}{\delta_{i}/[\lambda_{i}(1 - \delta_{i})]}^{1/2}$ ...where this
likelihood has been maximized by choice of the $\sigma^{**}$ estimate of $\sigma$ and by choice of
the positive and negative signs.

The $j^{th}-$component of the resulting ML shrinkage-estimator under normal-theory is:

\begin{equation}
\hat{\gamma}_j^{ML}=\hat{\delta}^{MSE}_jc_j=\biggl(\frac{n\hat{\rho}_j^2}{n\hat{\rho}_j^2%
+(1-R^2)}\biggr) \biggl(\hat{\rho}_j\sqrt{\frac{y^{\prime }y}{\lambda_j}}\biggr)\text{ ,}  \label{MLSE}
\end{equation}   % #14
\noindent where $y^{\prime}y = (n-1)$ when the response vector has been \textit{standardized}. Each
component, $\hat{\gamma}_j^{ML}$, of this ML estimator is a rational function of all $p$ principal
correlation estimators, $\hat{\rho}_1, \dots ,\hat{\rho}_p$. Note that the numerator consists of a single
term cubic in $\hat{\rho}_j$, while the denominator is quadratic in all of the principal correlation
estimators. In other words, this ML estimator is clearly \textit{not} a linear function of the
response $y-$vector.

Since linear models use \textit{conditional} distribution-theory where the $X-$matrix is
considered \textit{given}, rather than subject to random variation, the $H$ and $G$ matrices as
well as all functions of the eigenvalues of the $X-$matrix are also \textit{given} constants. The
$\hat{\gamma}_j^{ML}$ estimators of equation (\ref{MLSE}) are thus given functions of $y$ multiplied
by $\sqrt{y^{\prime }y/\lambda_j}$. In particular, note that the normal-theory conditional
distributions of the $\hat{\rho}_j-$estimators are \textit{not} those of correlation coefficients.

Unfortunately, the $\hat{\beta}(\Delta)$ estimate most likely to be $\hat{\beta}^{MSE}$ will \textbf{not}
always achieve smaller Summed MSE than OLS. After all, equation (\ref{MLSE}) shows that $\hat{\beta}(\Delta)$
is a \textit{non-linear} function of the $y-$vector.

The ``efficient path'' of Obenchain (2022a) consists of $p$ two-piece linear functions, each having a
single interior \textit{knot} at the $\hat{\beta}-$estimator with Maximum Likelihood of achieving minimum
MSE risk under normal distribution-theory. This new ``path'' is efficient in the senses that it is the shortest
path and, at least when $p > 2$, essentially the only known shrinkage path that always contains the
$\hat{\beta}-$vector that is most likely to be \textit{optimally biased}. Functions in R-packages freely
distributed via CRAN perform the calculations and produce graphics that illustrate optimal shrinkage. These
new concepts and visualization tools provide invaluable data-analytic insights and improved self-confidence
to applied researchers and data scientists fitting linear models to data.
 
Computations and Trace displays for the efficient GRR path (plus other graphics) were first
implemented in Version $2.0$ of the \textit{RXshrink} R-package, Obenchain (2022b).

Each GRR Trace typically displays estimates of $p$ quantities that change as \textit{shrinkage}
occurs. The ``coef'' Trace shows how fitted linear-model $\hat{\beta}-$estimates change with
shrinkage. The ``rmse'' Trace displays corresponding estimates of \textit{relative}
mean-squared-error given by dividing each diagonal element of the MSE-matrix by the OLS-estimate
of $\sigma^2$. When shrinkage becomes excessive, the ``infd'' Trace displays direction-cosine
estimates of the \textit{inferior-direction} in $p-$dimensional $X-$space along which over-shrunken
coefficients have higher MSE risk than their corresponding OLS estimates, Obenchain (1978).

The two final types of Trace diagnostics, called ``spat'' and ``exev'', are somewhat different;
they refer to the $p \geq 2$ rotated axes defining the \textit{principal-coordinates} of the
centered and rescaled $X-$predictors rather than to any single regression $\hat{\beta}-$coefficient
estimator. The additional restriction that $p \leq (n-4)$ then assures that $p{\times}p$ matrices
of unbiased estimates of MSE risk exist, Obenchain (1978, eq. (3.4)). However, to assure that
plotted relative risk values are at least as large as their relative variances, a
\textit{correct-range} estimate of relative MSE risk is displayed when the unbiased estimate would
be misleading.

While the above conventions have placed all $X-$information about the form and extent of any
\textit{ill-conditioning} into a convenient canonical-form, these conventions have done
nothing to predetermine the \textit{relative importance} of individual $x-$variables in
\textit{models} that predict $y-$outcomes. That information, as well as information on the
many effects of deliberate shrinkage, may well be best and most-clearly revealed via
\textit{visual} examination of Trace diagnostic plots.

\section{Quantifying Extent of Shrinkage} % Section 3...

A measure of the \textit{extent of shrinkage} applied by equation (\ref{BGRR}) is given by
\begin{equation}
m \equiv p - \delta_1 -\cdots- \delta_p = rank(X) - trace(\Delta).  \label{MCAL}
\end{equation}   % #15
\noindent This scalar, called the \textit{multicollinearity allowance}, Obenchain (1977),
is always $\geq 0$ and $\leq p =$ Rank of the $X-$matrix.

\vspace{0.5cm}
\textbf{Five KEY Advantages of using $m-$Scaling on ``Ridge TRACE'' Diagnostic Plots}
\begin{itemize}
\item {GENERALITY: Any sort of Shrinkage PATH can be displayed in a plot with \textbf{m} on its horizontal axis.}
\item {Rank DEFICIENCY: A vertical (dashed) line can be drawn on a TRACE display at the $m-$Extent
       most likely to represent the amount of $\delta-$factor shrinkage \textit{most likely to achieve
       minimum MSE risk}. When this $m-$value is strictly positive, it quantifies an approximate ``rank
       deficiency'' in the given $X-$matrix due to ill-conditioning (partial redundancy among columns).
       For examples, see the four TRACEs in Figure $4$ on page $15$.}
\item {FINITE Width and Height: All ``static'' $2-$dimensional plots have these restrictions, and $m \leq p$
       is clearly finite. }
\item {STABLE Relative Magnitudes: Shrunken regression coefficients and other estimates with
       \textit{perfectly stable relative magnitudes} form ``straight lines'' when plotted on TRACEs using
       $m-$scaling.}
\item {Bayesian Posterior Precision: For any given value of $m$, the average value of all $p$ shrinkage
       $\delta-$factors is $(p-m)/p$, which is also the proportion of Bayesian posterior precision due to
       \textit{sample information} ...rather than due to \textit{prior information}. This proportion thus
       decreases linearly as $m$ increases from $0$ to $p$.}
\end{itemize}
\vspace{0.5cm}
In the TRACE plot proposed by Hoerl and Kennard (1970a,b), their $k-$factor starts at $0$ but
must end abruptly at some finite $k-$max value specified by the user. Essentially, their shrinkage
terminus must be the ($0, ..., 0$) vector containing $p$ zeros, and $k-$max is apparently chosen
via trial-and-error. 

Use of $m-$scaling on the horizontal axis of Trace Diagnostic plots also suggests using
simplified notation, $\hat{\beta}_m$, to denote individual $\hat{\beta}(\Delta)$ estimators
in equation (\ref{BGRR}). The OLS solution ($\hat{\beta}^o$) which occurs at the beginning
($m=0$) of each GRR shrinkage-path would then be denoted by $\hat{\beta}_0$. Similarly,
$\hat{\beta}_p\equiv{0}$ occurs at the shrinkage terminus, $m = p$.

\section{Two-Parameter Shrinkage Paths}

Obenchain (1975) proposed restricted GRR Shrinkage Paths of the general form:

\begin{equation}
\delta_j = \lambda_j / (\lambda_j + k \times \lambda_j^q) = 1 / (1 + k \times \lambda_j^{(q-1)})\text{ .}  \label{2PAR}
\end{equation}

\noindent for $j = 1, ..., p$, where $k$ and $q$ are scalars such that $0 < k < +\infty$ and
$-5 \leq q \leq +5$. Goldstein and Smith (1974), equation (13), considered almost equivalent
paths except that their parameter m = 1 - q was assumed to be an integer.

The two well known special cases of equation (\ref{2PAR}) are (i) $q = 0$ for the ``ordinary'' path of
Hoerl and Kennerd (1970a,b) and (ii) $q = +1$ for Mayer-Willke (1973) ``uniform'' shrinkage,
$\hat{\beta}(\delta \cdot I) = \delta \cdot \hat{\beta}^o$.

\begin{figure}
\center{\includegraphics[width=0.5\textwidth]{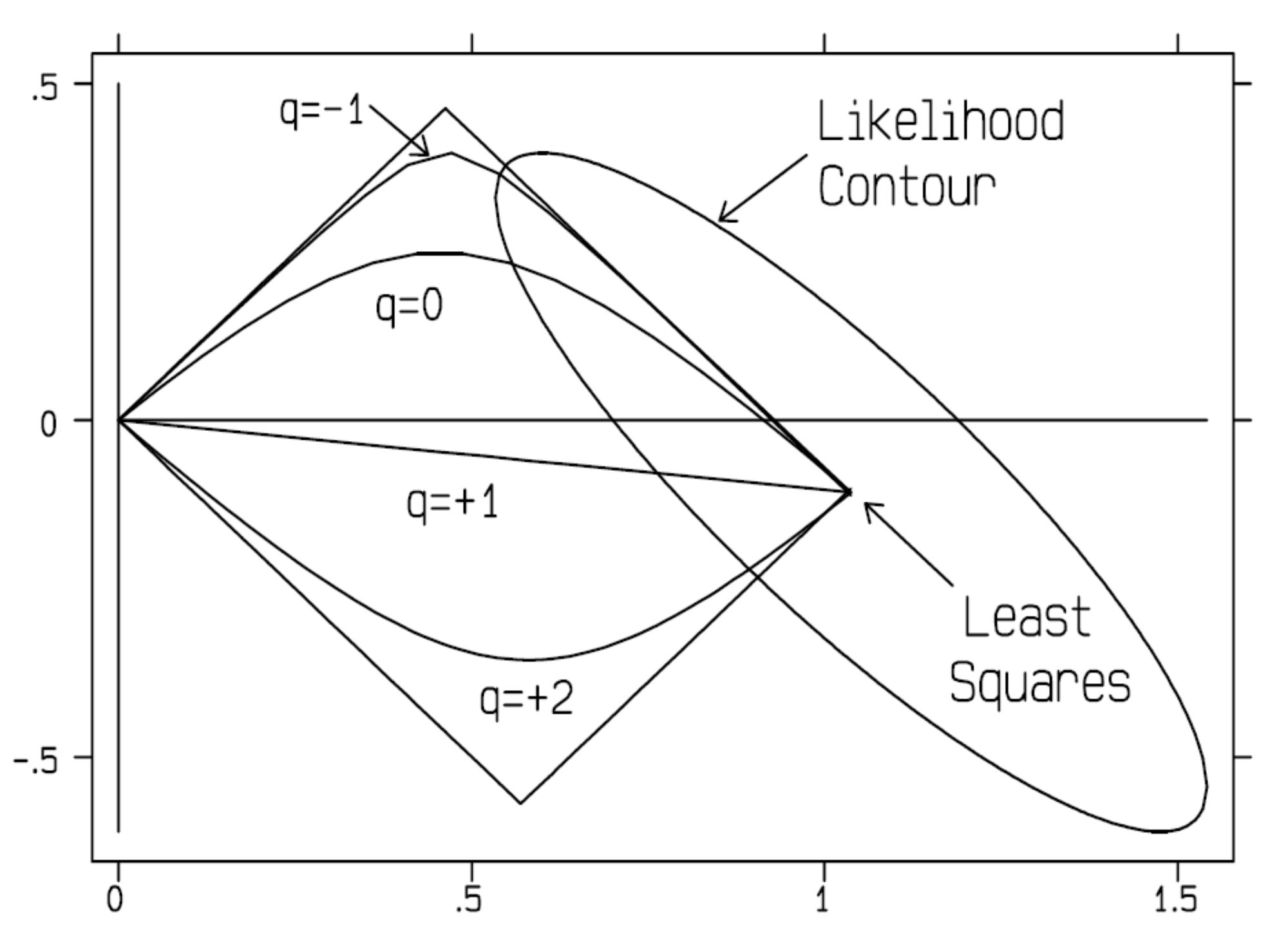}}
\caption{\label{fig:ridge2} Again, when the ``standardized'' $X-$matrix is of rank $p=2$,
the above plot shows paths of four different $q-$Shapes. Two issues strike me as intuitively clear
from this graphic. First, 2-parameter paths with $q < -1$ or $q > +2$ would be needed to come
closer to the upper and lower ``corners'', respectively, of the displayed shrinkage-factor rectangle.
Secondly, more parameters than just $k$ and $q$ are clearly needed when $p \geq 3$ to come anywhere
close to overall-optimal shrinkage, defined as $\delta_j \equiv \delta_j^{MSE}$ for $j = 1, ..., p$.}
\end{figure}

\subsection{Restricted Maximum Likelihood Shrinkage}

Note that equation (\ref{DOPT}) is easily solved to express the unknowns
($\gamma$ and $\sigma ^2$) as functions of the known eigenvalues
($\lambda _1,\cdots ,\lambda _r$) and of the MSE optimal shrinkage
factors ($\delta _1^{MSE},\cdots ,\delta _r^{MSE}$); this expression is 
\begin{equation}
{\gamma}_i=\pm \sigma \sqrt{\delta _i^{MSE}/[\lambda _i(1-\delta _i^{MSE})]}\text{ }.  
\label{GAMSIG}
\end{equation}
If an estimator within the restricted, 2-parameter shrinkage factor family
${\delta}_i=1/(1+k{\lambda}_i^{q-1})$ of equation (\ref{2PAR}) were MSE
optimal, equation (\ref{GAMSIG}) would then become

\begin{equation}
\gamma _i=\pm \sigma /\sqrt{k\lambda_i^q}.  \label{RESTKQ}
\end{equation}
Letting $\nu^2$ denote the common, unknown value of $\gamma _1^2\lambda
_1^q=\cdots =\gamma _r^2\lambda _r^q=\sigma ^2/k$ within our restricted
2-parameter search for MSE optimal shrinkage factors, the \textit{normal-theory
likelihood} that $\sigma ^2$ and $\gamma $ are of this highly restricted form
for any given values of $k$ (or $m$) and $q$ is $L(\gamma ,\sigma )=(2\pi
\sigma ^2)^{-n/2}e^{-u^2/2\sigma ^2}$ when $1^{\prime }y=0$ and
$1^{\prime }X=0^{\prime }$. The quadratic form in the exponential term
of  $L(\gamma ,\sigma )$ is then
\[u^2=(y-X\beta )^{\prime }(y-X\beta )\]
\begin{eqnarray}
=y^{\prime }y-2\sqrt{y^{\prime }y}\cdot \nu \cdot \sum |\rho_i|\lambda
^{(1-q)/2}+\nu^2\cdot \sum \lambda _i^{(1-q)} \label{QRATRHO}
\end{eqnarray}

\noindent because $y^{\prime }X\beta =y^{\prime }H\Lambda ^{1/2}G^{\prime }\beta =%
\sqrt{y^{\prime }y}\cdot \rho^{\prime}\Lambda ^{1/2}\gamma $ and $\gamma
_i=\pm \nu \lambda _i^{-q/2}$ under the restriction. Note, in particular,
that the numerical sign of each $\hat{\gamma}_i$ has been taken to agree
with its principal correlation, $\rho_i$, in equation (\ref{QRATRHO}); these
sign choices make the middle term of (\ref{QRATRHO}) as negative as
possible and reduce the quadratic form when $\nu > 0$ and p = rank($X$) is at
least $1$.

Once maximized by choice of a positive value of $\nu$, the
$L(\gamma ,\sigma )$ likelihood implied by equation (\ref{QRATRHO})
is, by definition, the likelihood that a given $k$ (or $m-$Extent) and $q-$Shape of
shrinkage yield MSE optimal $\delta-$factors.

Since the second derivative, $\partial ^2[u^2]/\partial{\nu^2}=2\cdot \sum
\lambda _i^{(1-q)}$, is strictly positive, the minimum of the $u^2$
quadratic form is achieved at $\partial [u^2]/\partial \nu = 0,$ which is
$\nu =\sqrt{y^{\prime}y}\cdot \sum |\rho_j|\lambda _j^{(1-q)/2}/\sum \lambda
_j^{(1-q)} > 0$ when r = rank($X$) is at least $1$. The corresponding minimum
$u^2$ is thus $u^2=y^{\prime }y\cdot [1-R^2CRL^2(q)],$ Obenchain(1975b),
for $R^2$ of (\ref{FRAT}) where the \textit{curlicue function} is

\begin{equation}
CRL(q)\equiv \frac{\sum \left| \rho_j\right| \lambda _j^{(1-q)/2}}{\sqrt{\sum
\rho_j^2\sum \lambda _j^{(1-q)}}}\text{ .}  \label{CURLICUE}
\end{equation}

\noindent Note that $CRL(q)$ is the Cosine of the angle between the R-vector of
absolute values of the \textit{principal correlations}
[$\rho$ of equation(\ref{UCORCOMP})]
and the L-vector of predictor eigenvalues raised to the power $(1-q)/2$.

\subsection{Optimal Choice of Path $q-$Shape}

Our crucial next step is to minimize the minus-two-log-likelihood of 
$n\cdot ln(2\pi \sigma ^2)+\hat{u}^2/\sigma ^2$ given $q$ by choice of a conditional
estimate of $\sigma ^2$, where $n$ is the number of observations.
Differentiating as usual, we find that the best choice for $\hat{\sigma}^2$ is
the minimum $\hat{u}^2$ divided by $n$. More importantly, the corresponding MSE
optimal $k-$factor, Obenchain(1981), can be written as
 
\begin{equation}
\hat{k}=\hat{k}(q)=\hat{\sigma}^2/\hat{\rho}^2=[\sum \lambda
_j^{(1-q)}]\cdot \frac{[1-R^2\cdot CRL^2(q)]}{n\cdot R^2\cdot CRL^2(q)}
\text{ .}  \label{KOPT}
\end{equation}

The final step is then to further minimize the $\hat{u}^2$ quadratic form
(maximize the likelihood) by maximizing $CRL(q)$ over choice of alternative
$q-$Shapes for the shrinkage path. While there is no known closed form
solution here, simple linear searches are fast within $-5\leq q \leq+5$.

Note that the arguments used here do not, technically, assume that each
$\sigma ^2/\gamma _i^2$ actually is equal to $k\lambda _i^q$. Rather, we are
simply asking: ``Which choice of $k$ and $q$ make it \textit{most likely} that the
unknown $\sigma ^2/\gamma _i^2$ are of this $k\lambda _i^q$ form
under normal distribution-theory?'' The corresponding
minus-twice-log-likelihood-ratio for the maximum likelihood 2-parameter
solution relative to the unrestricted solution of (\ref{DOPT}) is
 
\begin{equation}
\chi ^2(q)=n\cdot \ln \{1+\frac{R^2[1-CRL^2(q)]}{(1-R^2)}\}\text{ .}
\label{CHISQ}
\end{equation}

\noindent This large sample chi-squared test of the 2-parameter restriction has
$(r-2)$ degrees-of-freedom when $\chi ^2(q)$ has been minimized by choice
of $q-$Shape and r = rank($X$) $\geq 3$. A sufficiently large $\chi ^2(q)$ then
suggests that the 2-parameter family of (\ref{2PAR}) is too restrictive (smooth)
to contain the overall MSE optimal shrinkage ${\delta}-$factors.

The qm.ridge() function in the \textit{RXshrink} R-package of Obenchain(2022b) searches,
by default, over only integer and half-integer $q-$Shapes between qmin = $-5$ and qmax = $+5$.
The limit as $q$ approaches $+\infty$ is optimal for the Gibbons(1981) ``unfavorable'' case
where the true $\beta $ vector is parallel to the eigenvector with smallest eigenvalue,
$\lambda _p$. Shrinkage to $m=p-1$ ($\delta _1=\cdots =\delta _{p-1}=0$) then
reduces all components of $\hat{\beta}^o$ orthogonal to the true $\beta$ to zero! This is
essentially an extreme form of Massy(1965) ``type (b)'' principal components regression.

\begin{figure}
\center{\includegraphics[width=6in]{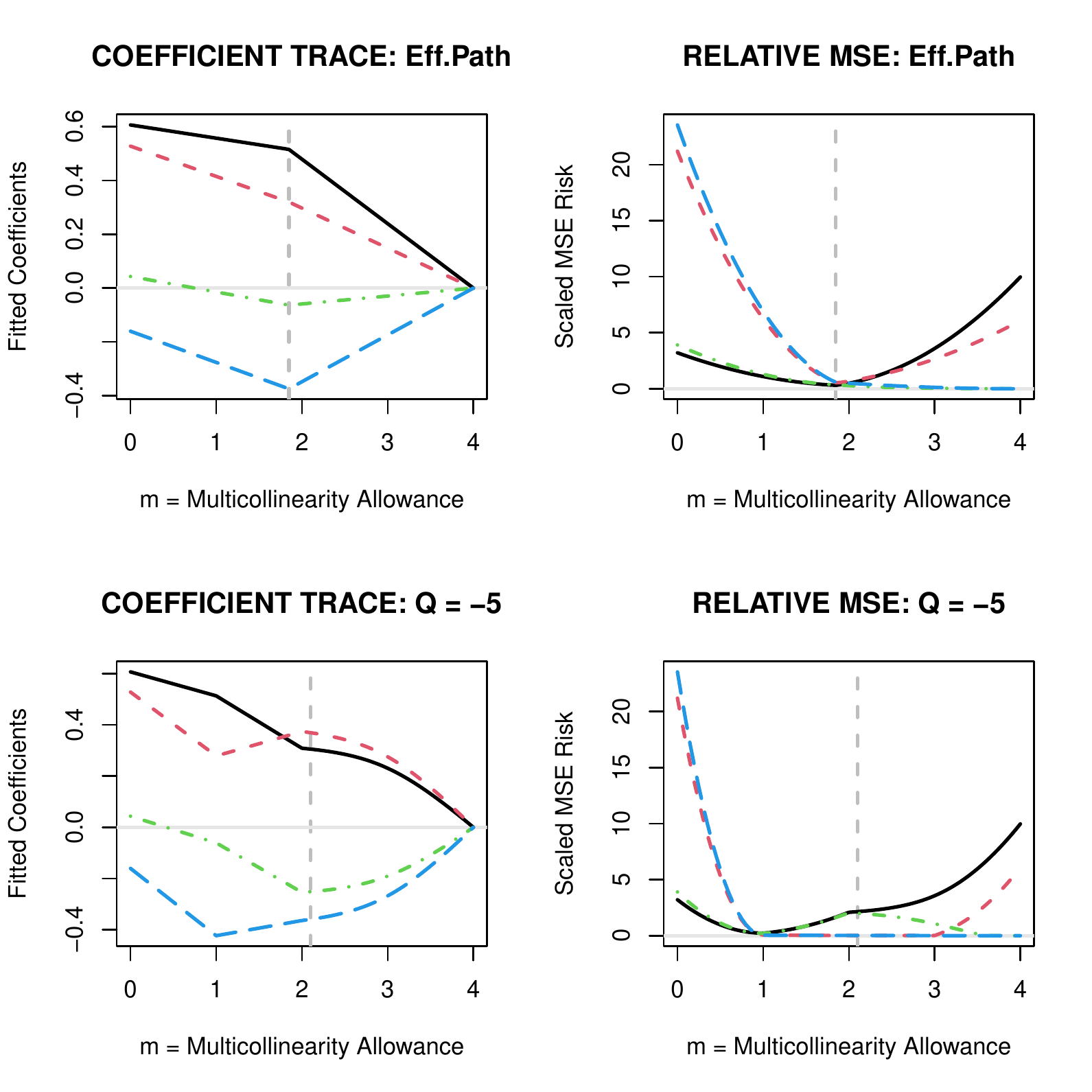}}
\caption{\label{fig:TRAC4} These are Efficient $4-$parameter and QM $2-$parameter TRACE
displays for the shrunken $\beta-$coefficients and their corresponding ``Relative'' MSE-error
estimates for the ``haldport'' dataset included in the RXshrink \textbf{R}$-$package. The vertical
dashed-lines indicating Maximum Likelihood of Minimal MSE Risk occur at $m = 1.85$ in the two top
panels and at $m = 2.12$ in the bottom panels. Note that the \textit{relative magnitudes} of ML
estimates (especially the solid-black and dashed-red coefficients) are quite different between these
two Paths.}
\end{figure}

\begin{figure}
\center{\includegraphics[width=6in]{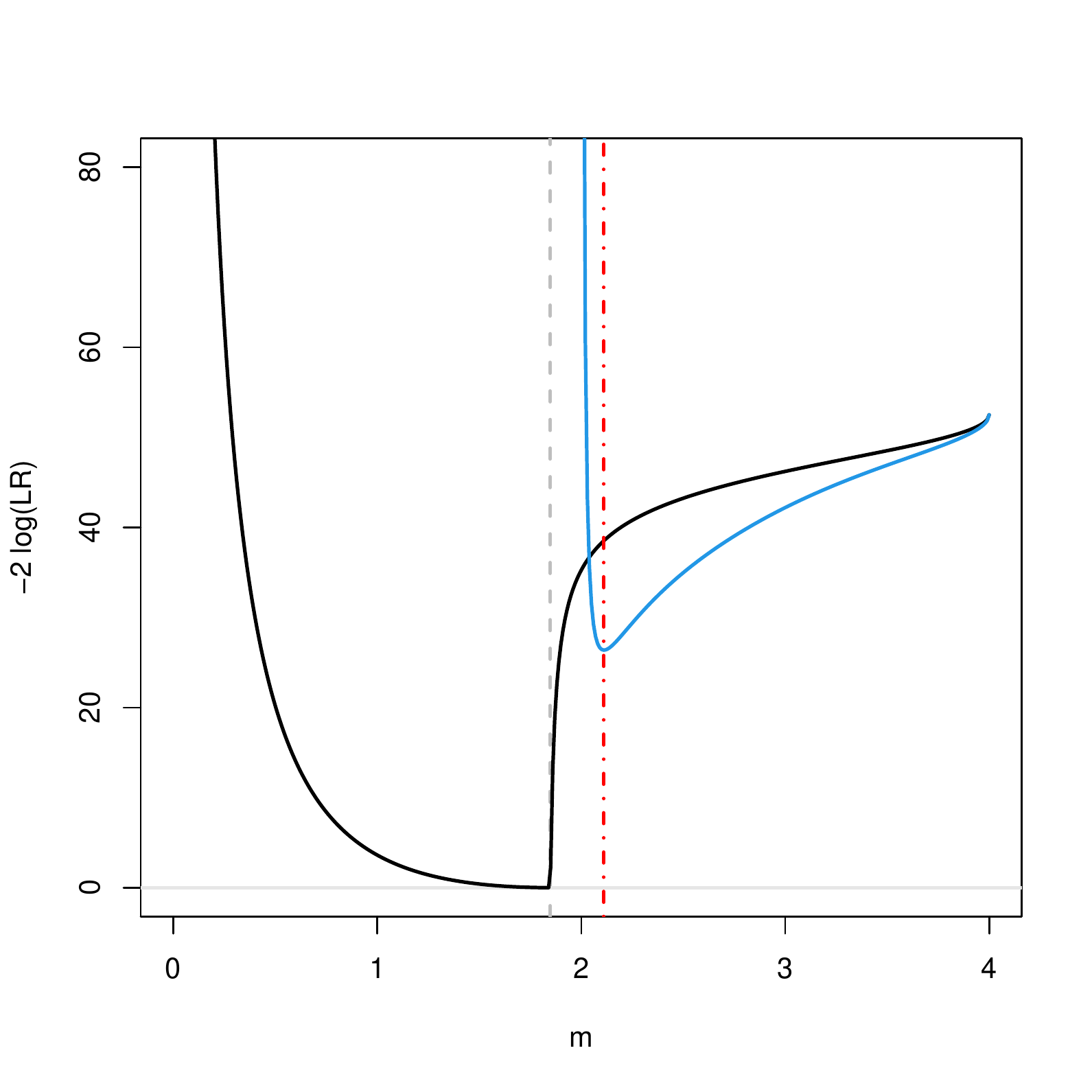}}
\caption{\label{fig:LRAT} Negative Log Likelihood Ratio curves for two different
shrinkage-paths on the ``haldport'' dataset. The solid black curve shows how the
$-2$ log Likelihood-Ratio under Normal-theory drops all of the way to $0$ at $m=1.85$
for the ``Efficient'' $4-$parameter path. The solid blue curve depicts the corresponding
L-R for the $2-$parameter path of ``best'' $q-\text{Shape}=-5$ on the default search grid.
This $-2 log(LR)$ reaches a minimum of $26.4$ at $m\approx{2.12}$. Since the upper $99\%-$point
of a $\chi^2-$variate with $2$ degrees-of-freedom is only $9.21$. Thus, the ``best''
$2-$parameter fit is actually significantly different from being MSE Risk-optimal.}
\end{figure}

Finally, the limit as $q$ approaches $-\infty$ contains all Massy(1965) ``type (a)''
principal components regression solutions. As in the above $q=+\infty$ case, the shrinkage
path travels along a series of edges of the generalized shrinkage hyper-rectangle. My
experience is that the $q=\pm 5$ paths are frequently adequate to approximate their
corresponding $q=\pm \infty$ limiting cases.

\section{Estimation of ``Relative'' MSE Risk}

An unbiased estimate of the ``scaled'' or ``relative'' Mean Squared Error matrix, $MSE(\Delta{c})/\sigma^2$, where
$\Delta$ is a non-stochastic diagonal matrix and $E(c) = \gamma$, is given by
\begin{equation}
\hat{T} = \frac{(n-p-3)}{(n-p-1)} (I - \Delta)\Lambda^{-1/2}\tau \tau{'}\Lambda^{-1/2}(I - \Delta)
+ \Lambda^{-1}{(2\Delta{-I})}\text{ ,}   \label{RELMSE}
\end{equation}
\noindent where $\tau$ is the column vector of t-statistics corresponding to the $F-$ratios of equation (\ref{FRAT}):
\begin{equation}
\tau_i = \rho_i \sqrt{\frac{n-p-1}{1-R^2}}\text{  for  }1 \leq i \leq p\text{.}     \label{TSTAT} 
\end{equation}
\noindent These $\tau_i-$estimators first appeared in equation ($3.4$) of Obenchain (1978) and are used
by functions in the RXshrink \textbf{R}$-$package, Obenchain (2022b), to display the ``Relative MSE''
Traces on the right-hand side of Figure(\ref{fig:TRAC4}).

When the $\delta_i$ are known and non-stochastic, the unknown relative risk, MSE($\delta_i{c_i})/\sigma^2$, must
be $\geq \delta_i^2/\lambda_i^2$ due to equation (\ref{MSE3}). Thus the known scaled variance of $\delta_i{c_i}$
is  $\delta_i^2/\lambda_i$, which provides a lower bound for the unknown relative MSE Risk of $\delta_i{c_i}$
when $\delta_i$ is non-stochastic.

\section{Simulated MSE Risk Comparisons}

The only published simulation results comparing the Maximum Likelihood approach described here with other methods
for choosing an extent of shrinkage are those of Gibbons (1981). She generated her ``O-method'' results
\textit{before} the closed-form expression, (\ref{KOPT}), for the maximum
likelihood $k-$factor was developed. Instead, Gibbons maximized
$CRL(q)$ over the range $-5\leq q\leq +1$ and then performed a
second search for the optimal shrinkage $k -$extent using the
``general'' likelihood monitoring equations of Obenchain(1975b). Gibbons
found that the O-method is superior to the Golub, Heath and
Wahba(1979) approach to generalized cross-validation (GHW) when $R^2$
exceeds 0.5 and the MSE optimal shrinkage pattern corresponds to $q = -\infty$
(i.e. the true $\beta $ vector lies along the first, major principal axis of
regressors.)

\begin{figure}
\center{\includegraphics[width=\textwidth]{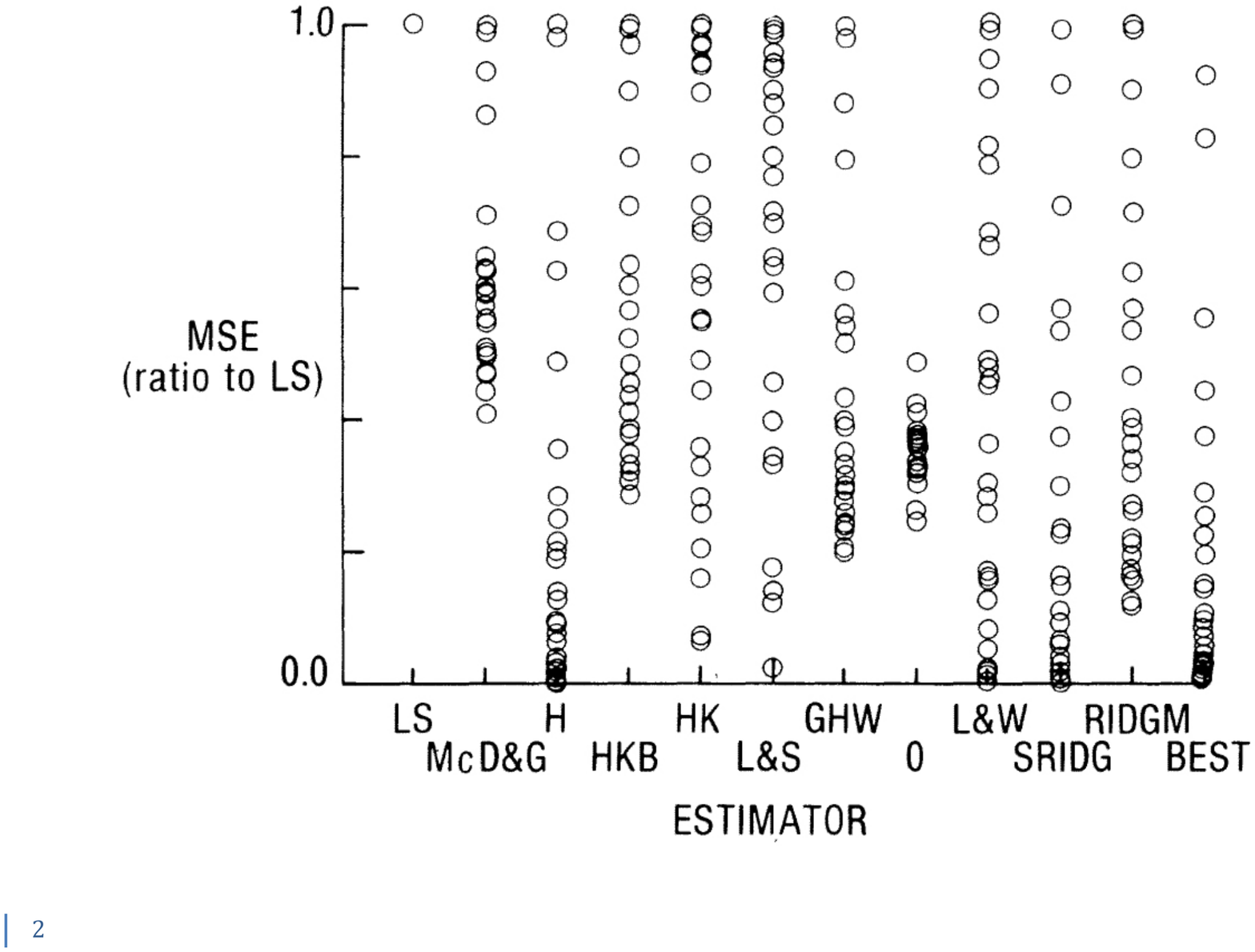}}
\caption{Here are some key comparisons from Figure 7 of Gibbons (1981). Note
that the O-method is conservative; it never shrinks aggressively enough to
reduce MSE risk to less than ~20\% of LS risk $=1.0$. Yet the maximum
risk of O-method selections is consistently less that 50\% of this LS risk.}
\label{fig:Gib7A}
\end{figure}

The top half of Figure 7 in Gibbons(1981), page 138, is reproduced here as
Figure \ref{fig:Gib7A}. Note in particular that, of the 12 methods simulated, only
the O-method consistently reduces MSE risk by at least 50\% in these so-called
``favorable'' cases.

The O-method could not perform well in Gibbon's ``unfavorable'' cases because
Gibbons never allowed the selected $q-$Shape to exceed $+1$, which corresponds to
uniform shrinkage, $\delta _1=\cdots =\delta _p=0$. However, Gibbons did report that
this most positive of allowed $q-$Shapes was, indeed, always selected by her ``O-method''
calculations in all simulations where the true $\beta$ vector lies along the last, minor
principal axis of regressors.

The 10 ``other'' shrinkage estimators that Gibbons was simulating (besides ordinary LS
and O) always follow the Hoerl-Kennard path of $q-$Shape = 0. This is a clearly
inappropriate overall pattern for shrinkage when the true $\beta$ vector is parallel to
the final principal direction of $p-$dimensional $X-$space with smallest, positive singular
value. This pathological case is (hopefully) quite rare in practical applications; after all,
why focus shrinkage on the final \textit{uncorrelated component}, $c_p$ of
(\ref{UCC2}), that contains the lone ``signal'' \textit{principal correlation}, $\rho_p$,
while minimizing shrinkage of the remaining pure ``noise'' components of the
ordinary LS solution? If the O-method had been allowed to select a $q-$Shape much
larger than $+1$, e.g. $+5$, in these Gibbons ``unfavorable'' scenarios, it would have
truly \textit{dominated} its 10 competitors on all measures of MSE risk.
 
The closed form expression, (\ref{KOPT}), for the maximum likelihood choice
of $k-$extent for shrinkage along any path of given $q-$Shape makes it
easier to study the MSE risk of O-method estimators using a variety of
techniques ...ranging from exact calculations to large sample approximations to
numerical integration to Monte-Carlo simulation. The primary challenge in
developing risk profiles for maximum likelihood shrinkage estimation lies
primarily in handling the high dimensionality of realistic scenarios. Specifically,
the extensive array of parameters that could be varied include
the relative sizes of true $\gamma $ components, spread in regressor
$\lambda $ eigenvalues, size of the error $\sigma ^2$ variance, number $p$ of
$X-$variables, singular $X-$matrices, number of degrees-of-freedom for error, etc, etc.

\section{Summary}

Over the last 50 years, computers have helped shape statistical theory as well as its practice.
Freely available software can provide computational and visual fast-tracks into the strengths and
weaknesses of alternative statistical methods. In fact, \textbf{R} and \textbf{Python} functions
are becoming almost indispensable tools for today's teachers of regression methods as well as for
students, applied researchers and data-scientists who use and/or extend these methods.

\end{document}